\documentclass[
 reprint,
 amsmath,amssymb,
aps,aps,twocolumn,superscriptaddress,showpacs,showkeys,amsmath,amssymb,floatfix
]{revtex4-2}
\usepackage[colorlinks=true,citecolor=blue,filecolor=blue,linkcolor=blue,urlcolor=blue,pdftex]{hyperref}

\usepackage{graphicx}
\usepackage{dcolumn}
\usepackage{bm}
\usepackage{placeins}
\usepackage[colorlinks=true,citecolor=blue,filecolor=blue,linkcolor=blue,urlcolor=blue,pdftex]{hyperref}
\RequirePackage{lineno}\newdimen\linenumbersep\linenumbersep=2pt
\usepackage{soul}

\newcommand{\bologna}{\affiliation{Department of Physics and Astronomy, University of Bologna and INFN-Bologna, 40126 Bologna, Italy}}
\newcommand{\chicago}{\affiliation{Department of Physics \& Kavli Institute for Cosmological Physics, University of Chicago, Chicago, IL 60637, USA}}
\newcommand{\coimbra}{\affiliation{LIBPhys, Department of Physics, University of Coimbra, 3004-516 Coimbra, Portugal}}
\newcommand{\columbia}{\affiliation{Physics Department, Columbia University, New York, NY 10027, USA}}
\newcommand{\lngs}{\affiliation{INFN-Laboratori Nazionali del Gran Sasso and Gran Sasso Science Institute, 67100 L'Aquila, Italy}}
\newcommand{\mainz}{\affiliation{Institut f\"ur Physik \& Exzellenzcluster PRISMA$^{+}$, Johannes Gutenberg-Universit\"at Mainz, 55099 Mainz, Germany}}
\newcommand{\mpik}{\affiliation{Max-Planck-Institut f\"ur Kernphysik, 69117 Heidelberg, Germany}}
\newcommand{\munster}{\affiliation{Institut f\"ur Kernphysik, Westf\"alische Wilhelms-Universit\"at M\"unster, 48149 M\"unster, Germany}}
\newcommand{\nikhef}{\affiliation{Nikhef and the University of Amsterdam, Science Park, 1098XG Amsterdam, Netherlands}}
\newcommand{\nyuad}{\affiliation{New York University Abu Dhabi - Center for Astro, Particle and Planetary Physics, Abu Dhabi, United Arab Emirates}}
\newcommand{\purdue}{\affiliation{Department of Physics and Astronomy, Purdue University, West Lafayette, IN 47907, USA}}
\newcommand{\rice}{\affiliation{Department of Physics and Astronomy, Rice University, Houston, TX 77005, USA}}
\newcommand{\stockholm}{\affiliation{Oskar Klein Centre, Department of Physics, Stockholm University, AlbaNova, Stockholm SE-10691, Sweden}}
\newcommand{\subatech}{\affiliation{SUBATECH, IMT Atlantique, CNRS/IN2P3, Universit\'e de Nantes, Nantes 44307, France}}
\newcommand{\torino}{\affiliation{INAF-Astrophysical Observatory of Torino, Department of Physics, University  of  Torino and  INFN-Torino,  10125  Torino,  Italy}}
\newcommand{\ucsd}{\affiliation{Department of Physics, University of California San Diego, La Jolla, CA 92093, USA}}
\newcommand{\wis}{\affiliation{Department of Particle Physics and Astrophysics, Weizmann Institute of Science, Rehovot 7610001, Israel}}
\newcommand{\zurich}{\affiliation{Physik-Institut, University of Z\"urich, 8057  Z\"urich, Switzerland}}
\newcommand{\paris}{\affiliation{LPNHE, Sorbonne Universit\'{e}, CNRS/IN2P3, 75005 Paris, France}}
\newcommand{\freiburg}{\affiliation{Physikalisches Institut, Universit\"at Freiburg, 79104 Freiburg, Germany}}
\newcommand{\napels}{\affiliation{Department of Physics ``Ettore Pancini'', University of Napoli and INFN-Napoli, 80126 Napoli, Italy}}
\newcommand{\nagoya}{\affiliation{Kobayashi-Maskawa Institute for the Origin of Particles and the Universe, and Institute for Space-Earth Environmental Research, Nagoya University, Furo-cho, Chikusa-ku, Nagoya, Aichi 464-8602, Japan}}
\newcommand{\laquila}{\affiliation{Department of Physics and Chemistry, University of L'Aquila, 67100 L'Aquila, Italy}}
\newcommand{\tokyo}{\affiliation{Kamioka Observatory, Institute for Cosmic Ray Research, and Kavli Institute for the Physics and Mathematics of the Universe (WPI), University of Tokyo, Higashi-Mozumi, Kamioka, Hida, Gifu 506-1205, Japan}}
\newcommand{\kobe}{\affiliation{Department of Physics, Kobe University, Kobe, Hyogo 657-8501, Japan}}

\newcommand{\kit}{\affiliation{Institute for Astroparticle Physics, Karlsruhe Institute of Technology, 76021 Karlsruhe, Germany}}
\newcommand{\tsinghua}{\affiliation{Department of Physics \& Center for High Energy Physics, Tsinghua University, Beijing 100084, China}}
\newcommand{\ferrara}{\affiliation{INFN - Ferrara and Dip. di Fisica e Scienze della Terra, Universit\`a di Ferrara, 44122 Ferrara, Italy}}

\begin{document}
\title{Search for events in XENON1T associated with Gravitational Waves}

\author{E.~Aprile}\columbia
\author{K.~Abe}\tokyo
\author{S.~Ahmed Maouloud}\paris
\author{L.~Althueser}\munster
\author{B.~Andrieu}\paris
\author{E.~Angelino}\torino
\author{J.~R.~Angevaare}\nikhef
\author{V.~C.~Antochi}\stockholm
\author{D.~Ant\'on Martin}\chicago
\author{F.~Arneodo}\nyuad
% ...

\collaboration{XENON Collaboration}
\email[]{rriya@purdue.edu}
\noaffiliation
%

\date{\today} 

\begin{abstract}
We perform a blind search for particle signals in the XENON1T dark matter detector that occur close in time to gravitational wave signals in the LIGO and Virgo observatories. No particle signal is observed in the nuclear recoil and electronic recoil channels within $\pm$\,500\,seconds of observations of the gravitational wave signals GW170104, GW170729, GW170817, GW170818, and GW170823. We use this null result to constrain mono-energetic neutrinos and axion-like particles emitted in the closest coalescence GW170817, a binary neutron star merger. We set new upper limits on the fluence (time-integrated flux) of coincident neutrinos down to 17~keV at 90\% confidence level. Furthermore, we constrain the product of coincident fluence and cross section of axion-like particles to be less than $10^{-29}$~cm$^2$/cm$^2$ in the [5.5--210]\,keV energy range at 90\% confidence level.
\end{abstract}


\maketitle

\section{\label{sec:level1}Introduction}
Following the first detection of a gravitational wave (GW) signal by the Laser Interferometer Gravitational-Wave Observatory (LIGO) in 2016~\cite{LIGOScientific:2016aoc}, GW signals from binary mergers are now routinely detected by the LIGO and Virgo observatories~\cite{LIGOScientific:2023vdi}. The subsequent observation of electromagnetic counterparts from binary neutron star coalescence in 2017~\cite{LIGOScientific:2017ync} has made GW signals an integral part of multi-messenger astrophysics. Various detectors have conducted follow-up searches for neutrinos associated with binary mergers in a time window of $\pm$\,$[400$--$500]$\,s of the observed GW signal~\cite{Santos:2019kdc, IceCube:2023atb, ANTARES:2023wcj, Super-Kamiokande:2018dbf, XMASS:2020jdj, KamLAND:2020ses, NOvA:2021zhv, BOREXINO:2023nji}, which might indicate the simultaneous emission of relativistic particles. The sensitive (neutrino) energy ranges of these searches, summarized in Tab.\,\ref{tab:Prev_search}, span from EeV down to half an~MeV. No evidence of a particle signal, coincident with GW signal, has been reported yet. 

\begin{table}[!htbp]
\caption{\label{tab:Prev_search} Search for particles coincident with GW and the energy ranges they are sensitive to.
}

\begin{ruledtabular}
\begin{tabular}{lrcl}
\hspace{0.25cm}\textrm{Detector}& \multicolumn{3}{c}{\hspace{-0.35cm}\textrm{Detection energy}}
\\
\colrule
\hspace{0.25cm}Pierre Auger Observatory~\cite{Santos:2019kdc} & \hspace{-0.2cm}10$^{11}$\,MeV &\hspace{-0.6cm}--& \hspace{-0.6cm}$10^{13}$\,MeV\rule{0pt}{2.5ex}\\
\hspace{0.25cm}IceCube~\cite{IceCube:2023atb} & \hspace{-0.2cm}10$^5$\,MeV &\hspace{-0.6cm}--& \hspace{-0.6cm}10$^{11}$\,MeV\\
\hspace{0.25cm}ANTARES~\cite{ANTARES:2023wcj} & \hspace{-0.2cm}10$^5$\,MeV &\hspace{-0.6cm}--& \hspace{-0.6cm}10$^{11}$\,MeV\\
\hspace{0.25cm}Super-K~\cite{Super-Kamiokande:2018dbf} & \hspace{-0.2cm}3.5\,MeV &\hspace{-0.6cm}--& \hspace{-0.6cm}10$^{11}$\,MeV\\
\hspace{0.25cm}XMASS~\cite{XMASS:2020jdj} & \hspace{-0.2cm}14\,MeV &\hspace{-0.6cm}--& \hspace{-0.6cm}100\,MeV\\
\hspace{0.25cm}KamLAND~\cite{KamLAND:2020ses} & \hspace{-0.2cm}1.8\,MeV &\hspace{-0.6cm}--& \hspace{-0.6cm}111\,MeV\\
\hspace{0.25cm}NOvA~\cite{NOvA:2021zhv} & \hspace{-0.2cm}10\,MeV &\hspace{-0.6cm}--& \hspace{-0.6cm}100\,MeV\\
\hspace{0.25cm}Borexino~\cite{BOREXINO:2023nji} & \hspace{-0.2cm}0.5\,MeV &\hspace{-0.6cm}--& \hspace{-0.6cm}5\,MeV\\
\hspace{0.25cm}XENON1T (this work) & \hspace{-0.2cm}0.017\,MeV &\hspace{-0.6cm}--& \hspace{-0.6cm}50\,MeV\\
\end{tabular}
\end{ruledtabular}
\end{table}

Here, we use data from the XENON1T dark matter search~\cite{XENON:2017lvq} to extend this neutrino energy window down to 17~keV. Further, the low threshold of 1~keV electron-equivalent (keV$_{\mathrm{ee}}$) offered by the XENON1T, enables us to search for axion-like particles (ALPs) in the keV energy range. There have been studies proposing the emission of relativistic axions~\cite{Harris:2020qim, Fiorillo:2022piv} from the merger of neutron stars. Since blackbody radiation peaks at several to hundreds of keV for binary neutron stars~\cite{Wang:2018xhm}, ALPs might exist in thermal equilibrium with the blackbody radiation with keV energy. Additionally, ALPs can be accelerated to keV energies by 511-keV line associated with positrons produced by mergers~\cite{Fuller:2018ttb}. Not only the low energy thresholds of XENON1T are helpful in looking for these keV ALPs, but they also allow looking for MeV Beyond Standard Model particles (like sterile neutrinos~\cite{Frensel:2017prj, Frensel:2016fge}) which deposit only a fraction of their energy via interactions like coherent scattering~\cite{Abdullah:2022zue}.

\section{Data sets}\label{datasets}

\begin{table*}[!htbp]
\caption{\label{tab:Channels} Characteristics of analyses considered in this study. For each of these analyses, we show the corresponding energy thresholds, the Science Runs that have been analyzed for that channel, the fiducial target mass, and the observed background rate. Further, in the $\pm$\,500\,s time window around a GW event, we give the average livetime for detecting a signal, the average number of expected background events (N$_{\mathrm{exp}}$), and the probability of observing zero background events (P(0)).}
\begin{ruledtabular}
\begin{tabular}{
			>{\raggedright}m{2cm} 
			>{\raggedleft}m{1.6cm}
			c
			>{\raggedright}m{1.5cm} 
			>{\centering}m{1.6cm} 
			>{\centering}m{1.7cm}
            >{\centering}m{2.4cm} 
			>{\centering}m{1.4cm} 
			>{\centering}m{1.8cm} 
			>{\centering\arraybackslash}m{1.8cm} 
		}
 & \multicolumn{3}{c}{Energy} & Runs & {Fiducial} & Background rate&   Livetime& \\ 
Analyses & \multicolumn{3}{c}{threshold} & analyzed & {target mass} &(events/day)&  (per GW)& N$_{\mathrm{exp}}$ & P(0)\\ \hline
ER~\cite{XENON:2020rca} & \multicolumn{3}{c}{1\,keV$_{\mathrm{ee}}$}  & SR1 & 1.04\,tonne& 186 & 935\,s &  See Tab.\,\ref{tab:ER_Table} & See Tab.\,\ref{tab:ER_Table}\rule{0pt}{2.5ex}\\
3-fold~\cite{XENON:2017vdw, XENON:2018voc}  & \multicolumn{3}{c}{4.9\,keV$_{\mathrm{nr}}$} & SR0, SR1 & 1.3\,tonne& 0.05 & 935\,s & 0.00054 & 0.9995\\
2-fold~\cite{XENON:2020gfr} & \multicolumn{3}{c}{1.6\,keV$_{\mathrm{nr}}$} & SR1 & 1.04\,tonne& 0.03 & 790\,s & 0.00027 & 0.9997\\
S2-only~\cite{XENON:2019gfn} & \multicolumn{3}{c}{0.7\,keV$_{\mathrm{nr}}$} & SR1 & 0.12\,tonne& 0.94 & 650\,s & 0.0071 & 0.993\\
\end{tabular}
\end{ruledtabular}
\end{table*}

The LIGO and Virgo observatories conducted their second observation run from November 30, 2016, to August 25, 2017~\cite{LIGOScientific:2018mvr}. Overlapping with this time frame, the XENON1T detector took Science Run 0 (SR0) data from November 22, 2016 to January 18, 2017, and Science Run 1 (SR1) data from February 2, 2017 to February 8, 2018. This data has already been analyzed in several modes that we revisit here in order to check for particle signals coincident with LIGO and Virgo GW observations. 

The XENON1T experiment has been described in~\cite{XENON:2017lvq}. There are two channels via which a particle can interact with the detector. A particle can scatter off the xenon nuclei, referred to as nuclear recoil (NR). Similarly, a particle can scatter off the electrons of xenon atoms, referred to as electronic recoil (ER)~\cite{XENON:2020rca}. There are three ways in which searches for NR signals have been performed. The search for weakly interacting massive particles has been done demanding both scintillation light (S1) and electroluminescence from ionization electrons (S2)~\cite{XENON:2017vdw, XENON:2018voc}. This will be referred to as 3-fold in the rest of the paper because coincident signals from at least three photomultiplier tubes (PMTs) are required for an S1. However, with the purpose of decreasing the energy threshold, the data have been analyzed in two other modes. To detect NRs produced by solar neutrinos via coherent elastic neutrino-nucleus scattering (CE$\nu$NS), the requirement for an S1 has been loosened to coincident signals from at least two PMTs~\cite{XENON:2020gfr}. This will be referred to as 2-fold in the rest of the paper. To detect NRs produced by light dark matter, the requirement of a scintillation signal has been waived~\cite{XENON:2019gfn}, referred to as S2-only in the rest of the paper. 

The 3-fold search for NRs was done using data from both SR0 and SR1. However, the search for ERs or 2-fold and S2-only searches for NRs were done only using data from SR1. Details of these analyses can be found in the respective papers~\cite{XENON:2017vdw, XENON:2018voc, XENON:2020rca, XENON:2020gfr, XENON:2019gfn}. To re-iterate, the search for ERs and the 3-fold search for NRs demand coincident signals from at least three PMTs for an S1, while the 2-fold requires signals from only two PMTs, and S2-only doesn't require S1. Furthermore, the S2 threshold has been reduced from 200 photoelectrons (PE) for 3-fold analysis to 120\,PE and 150\,PE for 2-fold and S2-only analyses, respectively. The S2 threshold for ER channel is 500 PE. The corresponding energy thresholds for these analyses are given in Tab.\,\ref{tab:Channels} in units of keV electron-equivalent (keV$_{\mathrm{ee}}$) for ER channel, and as keV nuclear recoil-equivalent (keV$_{\mathrm{nr}}$) for NR channel. 

The lower threshold for 2-fold and S2-only analyses comes with the cost of increased backgrounds, to reduce which, stringent cuts need to be applied. For 2-fold analysis, these cuts result in low total acceptance -- peak acceptance is $\sim$15\%. The acceptance is lower than 10\% for energy region outside 2-4\,keV$_{\mathrm{nr}}$, becoming negligible above 6\,keV$_{\mathrm{nr}}$; compared to $\sim$80\% acceptance for 3-fold and S2-only analyses within 10-40\,keV$_{\mathrm{nr}}$ and 2-20\,keV$_{\mathrm{nr}}$ respectively. On the other hand, the stringent radius and width cuts of S2-only analysis lead to the fiducial target mass of 0.12 tonne compared to fiducial target mass of $>$\,1\,tonne in 2-fold and 3-fold analyses.  

The background rate is very low for NR channel (see Tab.~\ref{tab:Channels}), which motivates using a single energy bin for each analysis mode. However, the background rate is quite high in ER channel due to the $\beta$ decay of $^{214}$Pb, present due to $^{222}$Rn emanation by materials~\cite{XENON:2020rca}. Further, from Fig.\,3 of~\cite{XENON:2020rca}, we can see that the ER spectrum is not constant in energy. The spectrum exhibits spectral lines from radioactive $^{83\mathrm{m}}$Kr and $^{131\mathrm{m}}$Xe. We thus define five energy bins for the ER channel: low energy (0--30\,keV$_{\mathrm{ee}}$), $^{83\mathrm{m}}$Kr (30--50\,keV$_{\mathrm{ee}}$), medium energy (50--142\,keV$_{\mathrm{ee}}$), $^{131\mathrm{m}}$Xe (142--185\,keV$_{\mathrm{ee}}$) and high energy (185--210\,keV$_{\mathrm{ee}}$). 

Three of the eight GW signals that LIGO/VIRGO observed during SR0 and SR1 occurred during XENON1T calibration or downtime. GW170104 was during SR0 and therefore only 3-fold analysis is available for it. GW170729, GW170817, GW170818, and GW170823 were during SR1, and so all four analyses are available for them. Out of these GWs, GW170817 was due to the merger of neutron stars and the only GW event with a co-detection of electromagnetic signatures. Further, it was at a distance of 40~Mpc, compared to a distance of $\sim$1000~Mpc for other events. While the particle flux on the Earth would be influenced by initial masses, nuclear equation of state, and anisotropies of emission~\cite{Cusinato:2021zin}, the proximity to merger also plays a role because the flux is inversely proportional to the square of the distance. Because of all these reasons, GW170817 is the most interesting and would be used to derive our results in section~\ref{results}.

\section{Neutrino spectra}

Neutrinos predominantly interact in our detector via elastic scattering off electrons or xenon nuclei. The standard model differential cross section for elastic neutrino-electron scattering can be written as~\cite{Giunti:2008ve, Vogel:1989iv},

\begin{equation}\label{cross-section} 
\begin{split}
 \frac{d\sigma_\mathrm{ES}}{dE_r} = \frac{G_F^2m_e}{2\pi}&\left[(g_V + g_A)^2 -(g_V^2-g_A^2)\frac{m_eE_r}{E_\nu^2}  \right.\\
     &\left.+ (g_V-g_A)^2\left(1-\frac{E_r}{E_\nu}\right)^2\right],    
\end{split}
\end{equation}
where $E_r$ is the ER kinetic energy, $E_\nu$ is the incoming neutrino energy, $m_e$ is the mass of an electron, and $G_F$ is the Fermi constant. Neutrino-electron coupling constants for different flavors of neutrinos and antineutrinos are furnished in Tab.\,\ref{tab:coupling}.

\begin{table}[!htbp]
\caption{\label{tab:coupling} Neutrino-electron coupling constants for different flavors of neutrinos and antineutrinos. $\theta_W$ is the Weinberg angle.}
\begin{ruledtabular}
\begin{center}
    \begin{tabular}{ccc}
Neutrino flavor&$g_V$&$g_A$\\
\colrule
$\nu_e$&$2\,\mathrm{sin}^2\theta_W + \frac{1}{2}$&$+\frac{1}{2}$\rule{0pt}{3ex}    \\
$\bar{\nu}_e$&$2\,\mathrm{sin}^2\theta_W + \frac{1}{2}$&$-\frac{1}{2}$\rule{0pt}{3ex}    \\
$\nu_\mu$, $\nu_\tau$&$2\,\mathrm{sin}^2\theta_W - \frac{1}{2}$&$-\frac{1}{2}$\rule{0pt}{3ex}\\
$\bar{\nu}_\mu$, $\bar{\nu}_\tau$&$2\,\mathrm{sin}^2\theta_W - \frac{1}{2}$&$+\frac{1}{2}$\rule{0pt}{3ex}
\end{tabular}
\end{center}
\end{ruledtabular}
\end{table}

The expected number of signals in the ER channel is given as, 
\begin{equation} \label{spectra equation}
    \frac{dN_\mathrm{ER}}{dE_r} = N_T(E_r) \epsilon(E_r)\int \frac{d\sigma_\mathrm{ES}}{dE_r}\frac{df(E_\nu)}{dE_\nu}dE_\nu,
\end{equation}
where $\epsilon(E_r)$ is the detection efficiency of the ER channel as a function of the recoil energy~\cite{XENON:2020rca}. $f(E_\nu)$ is the neutrino energy spectrum. Given an absence of theoretical predictions, we assume that neutrinos are emitted mono-energetically in binary mergers, as done in previous low-energy searches~\cite{Super-Kamiokande:2018dbf, XMASS:2020jdj, BOREXINO:2023nji}. $N_T(E_r)$ is the total number of available electrons in the fiducial target. The atomic binding suppresses the cross sections by $\sim$20\%, as shown in Fig. 2 of ~\cite{Chen:2016eab}, using relativistic random phase approximation (RRPA). Since cross section values from RRPA are available only till 30 keV$_{\mathrm{ee}}$, while our energy region of interest extends to 210 keV$_{\mathrm{ee}}$; we use the stepping approximation --- electrons in atomic shells with binding energy lower than the recoil energy are considered to be free. Therefore, $N_T(E_r) = N_{\mathrm{Xe}} \times N_\mathrm{e}$, where $N_\mathrm{Xe}$ is the total number of xenon atoms in the fiducial volume and $N_\mathrm{e}$ is number of electrons per xenon atom with binding energies below the particular recoil energy. The different shells of a xenon atom, along with the number of electrons and the potential energy, are given in~\cite{winter}. We also convolve the cross section with a Gaussian function with standard deviation $0.31 \sqrt{E}\sqrt{\mathrm{keV}} + 0.0037 E$~\cite{XENON:2020rca} to account for energy resolution, where E is in keV.


\begin{figure*}[!htbp]
\includegraphics[scale =0.45]{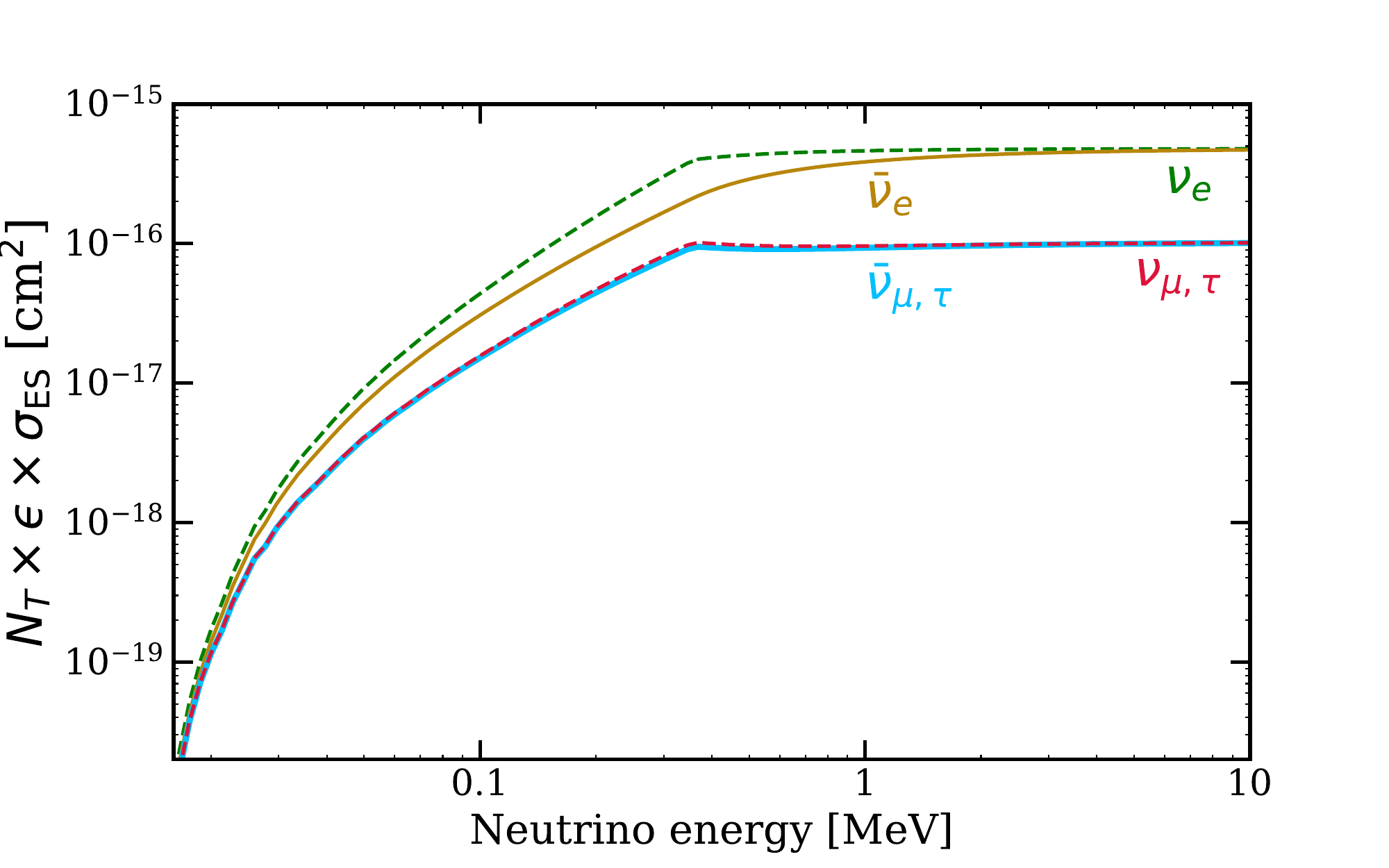}\includegraphics[scale =0.45]{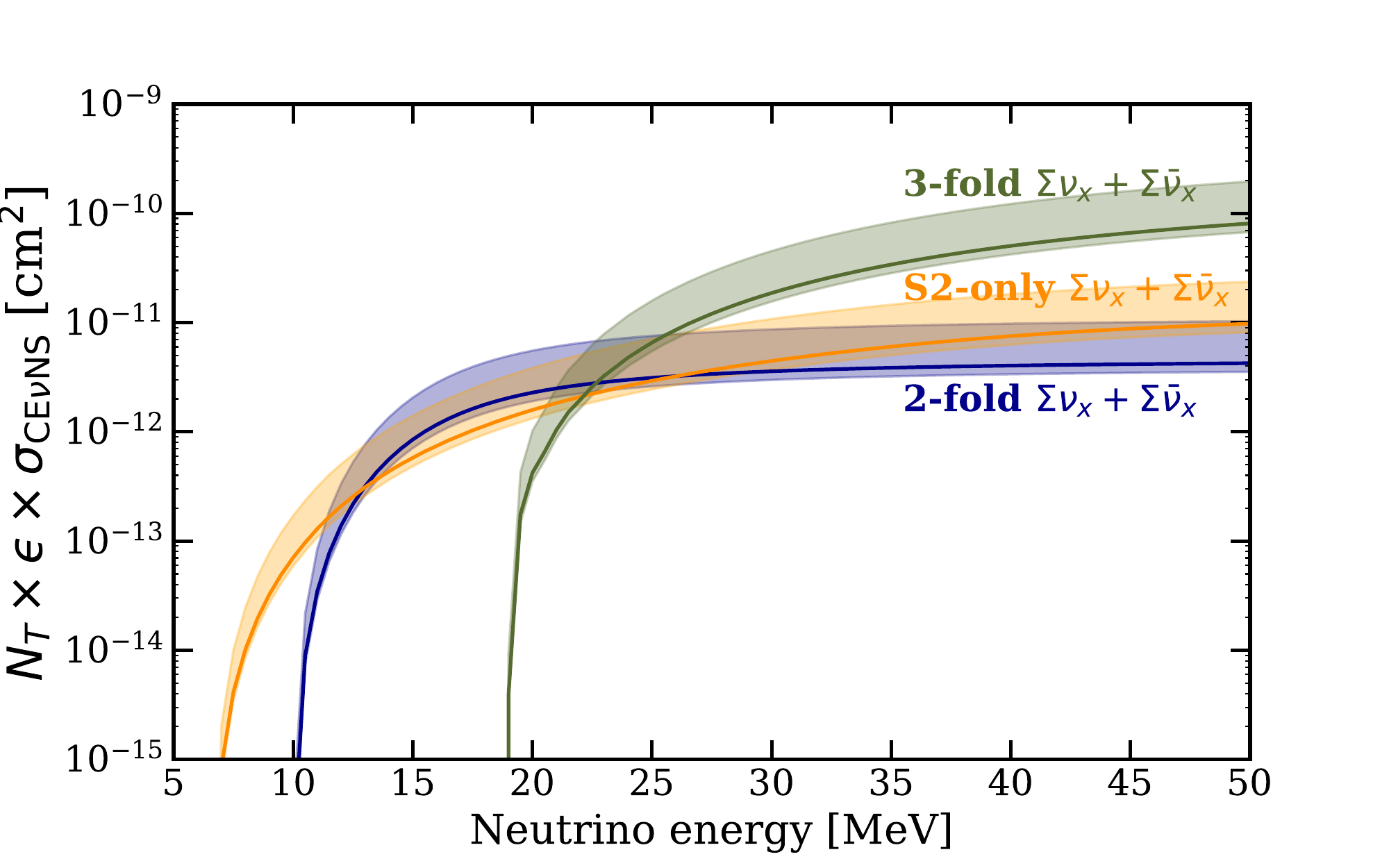}
\caption{\label{Neutrino_CC} Left: the effective cross section for elastic neutrino-electron scattering as a function of neutrino energy for electron neutrinos ($\nu_e$, dashed green), electron antineutrinos ($\bar{\nu}_e$, solid gold), muon or tau neutrinos (${\nu}_{\mu, \tau}$, dashed red), and muon or tau antineutrinos ($\bar{\nu}_{\mu, \tau}$, solid cyan). Right: the effective cross section for CE$\nu$NS as a function of neutrino energy. 3-fold (olive-green), 2-fold (blue), and S2-only (orange) refer to different modes in which the data was analyzed in the search for nuclear recoils, as described in Sec.\,\ref{datasets}. In the right plot, the cross section is for the interaction of all flavors of neutrinos and antineutrinos.}
\end{figure*}

The effective cross section as a function of neutrino energy, taking into account the detector properties, is shown in Fig.\,\ref{Neutrino_CC} (left). As can be seen in equation~\ref{cross-section}, neutrino energy is a higher-order term for differential cross section. The change in the physical cross section of electron neutrinos, electron antineutrinos, muon or tau neutrinos, and muon or tau antineutrinos with neutrino energy is only 16\%, 54\%, 1.2\%, and 5\% respectively. However, to satisfy the kinematics of the system, a neutrino can not produce a recoil of energy greater than $2 E_\nu^2 / (2E_\nu + m_e)$~\cite{Vogel:1989iv}, i.e., the cross-section sharply goes to 0 at this recoil energy for a given neutrino energy. Our upper threshold for ER energy is 210~keV$_{\mathrm{ee}}$, which is the maximum ER energy produced by a 360 keV neutrino. Lower energy neutrinos are incapable of generating ERs throughout the entire energy range of the ER channel (1\,keV$_{\mathrm{ee}}$ -- 210\,keV$_{\mathrm{ee}}$). This leads to a drastic decrease in the effective cross section below 360~keV. Further, our lower threshold of 1~keV$_{\mathrm{ee}}$ makes us insensitive to neutrinos of energy 16~keV or below.

Having discussed the interaction of neutrinos with electrons, we now turn to the interaction of neutrinos with xenon nuclei. Neutrinos in the energy range of 10 -- 50 MeV can interact coherently with nuclei~\cite{Abdullah:2022zue}. For CE$\nu$NS, we use the same differential cross section as in~\cite{XENON:2020gfr}. The expected number of signals in the NR channel has a similar form as Eq.~(\ref{spectra equation}), with $N_T(E_r)$ and $\sigma_\mathrm{ES}$ replaced by $N_{\mathrm{Xe}}$ and $\sigma_{\mathrm{CE}\nu\mathrm{NS}}$ respectively. Also, $\epsilon(E_r)$ is the detection efficiency of considered analysis mode as a function of the NR kinetic energy. Assuming mono-energetic emission of the neutrinos, the effective cross section as a function of neutrino energy, taking into account the detector properties, is shown in Fig.\,\ref{Neutrino_CC}\,(right). Effective cross section is highest for 3-fold analysis because of larger acceptance and fiducial mass. The lower fiducial mass for S2-only analysis leads to an order of magnitude lower effective cross section at higher neutrino energy. The very low total acceptance of 2-fold analysis leads to even lower effective cross section. However, because of comparatively higher acceptance at lower recoil energies, the 2-fold effective cross section is comparable or even larger than S2-only effective cross section at lower neutrino energies.

We also consider non-standard neutrino-nuclei interactions via the vector coupling of $\nu_\mathrm{e}$ to up and down quarks. 
The coupling values within 90\% confidence interval from XENON1T data \cite{XENON:2020gfr}, can effectively change the total interaction cross section from a factor of 0.83 to 2.42, under the assumption that 1/6$^{th}$ of the total neutrinos are electron neutrinos. The resulting changes in the number of NRs are shown as an uncertainty band in Fig.\,\ref{Neutrino_CC}\,(right). 

\section{Background Estimation}

Consistent with previous coincidence searches shown in Tab.\,\ref{tab:Prev_search}, which used a time window of $\pm$\,$[400$--$500]$\,s, we use a $\pm$\,$500$\,s time window centered on the time of GW signal. This is appropriate for any relativistic particle flux emitted by the binary merger~\cite{Baret:2011tk}. However, in XENON1T, various cuts in different analyses remove events at certain times, accounted for as deadtime~\cite{XENON:2019ykp}. These deadtimes are uniformly distributed throughout the 1000 seconds time windows. We remove 5 milliseconds around a trigger from the muon veto ~\cite{XENON1T:2014eqx} to reject muons or muon-induced showers passing through the detector. There are 323--390 such triggers within 1000 seconds of gravitational wave events, resulting in a deadtime of $\sim1.8$\,s. There are also veto cuts to remove events if the data acquisition system (DAQ) is unable to record the data. This occurred 96--228 times in these 1000 second windows, resulting in a deadtime of $[2.6$--$16.7]$\,s. Further, event duration is defined by the maximum drift time in the detector, which is 750\,$\mu s$, resulting in $\sim14$\,s deadtime due to background events.

There are also cuts to remove lone S1s and S2s produced by energy deposition in light- and charge-insensitive regions. These cuts are tightened for 2-fold analysis because of the higher background rate, resulting in its lower livetime. We multiply the remaining livetime with the cut-acceptance of these cuts to account for deadtime due to the cuts. The livetimes obtained for detecting a signal in the $\pm$\,$500$\,s time window around a GW signal in the ER, 3-fold, 2-fold and S2-only analyses are $[926$--$942]$\,s, $[926$--$942]$\,s, $[785$--$792]$\,s and $[920$--$933]$\,s respectively. For S2-only analysis, 30\% of events, uniformly distributed over time, was taken to train the cuts. Therefore, the livetime of S2-only analysis further reduces by 30\% to $[644$--$653]$\,s. The average livetimes for different analyses are given in Tab~\ref{datasets}.

\begin{figure*}[!htbp]
\includegraphics[scale =0.33]{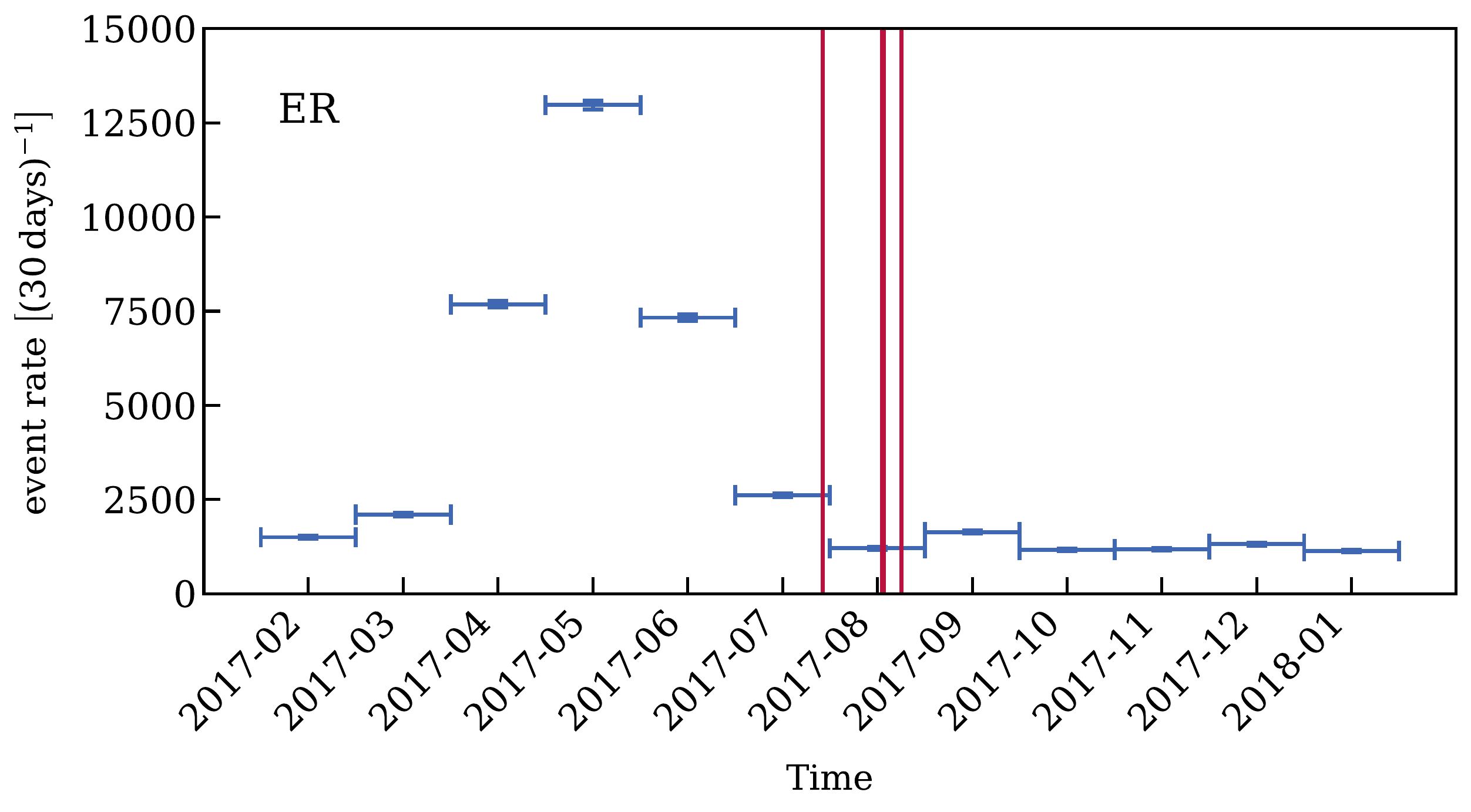}
\includegraphics[scale =0.33]{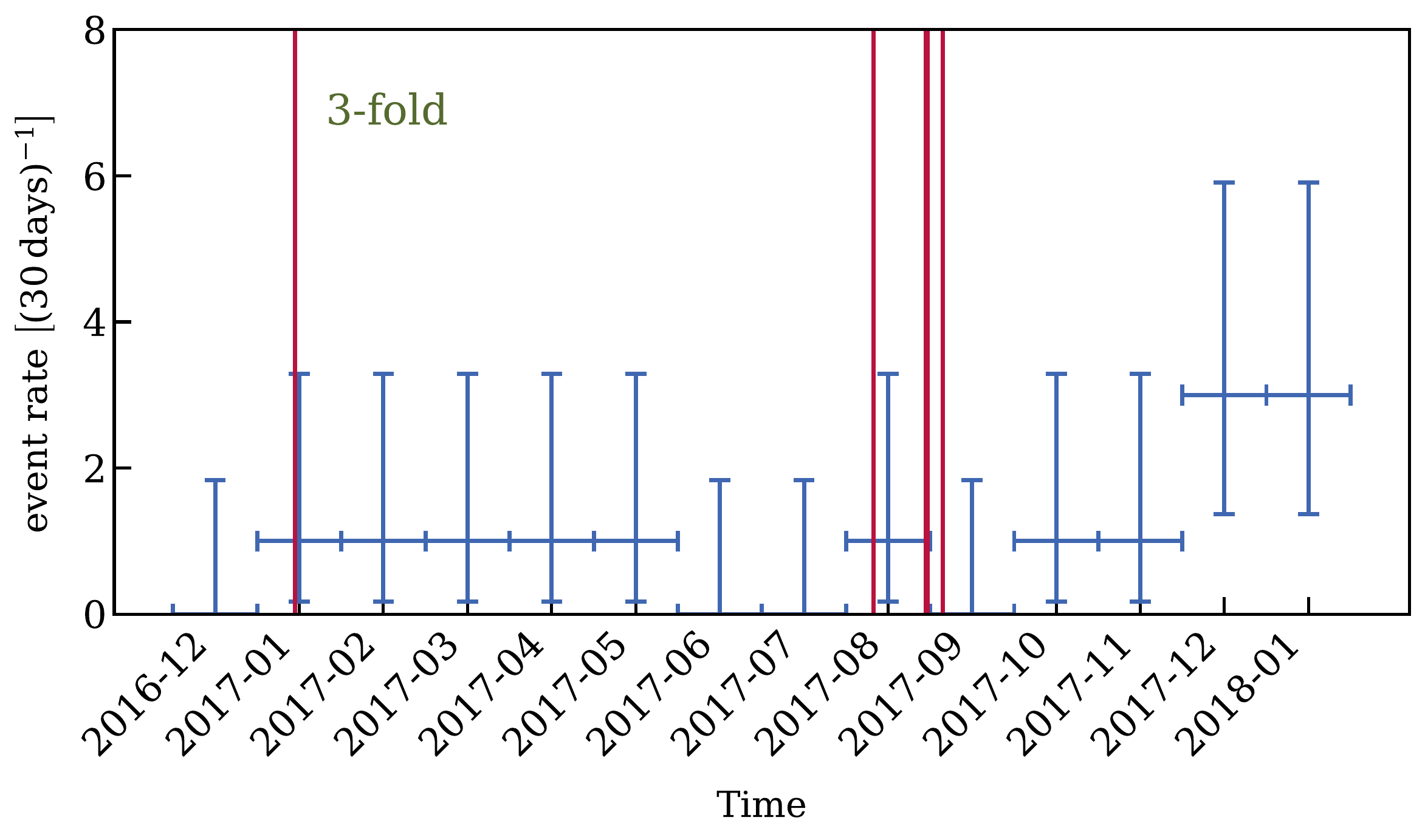}
\includegraphics[scale =0.34]{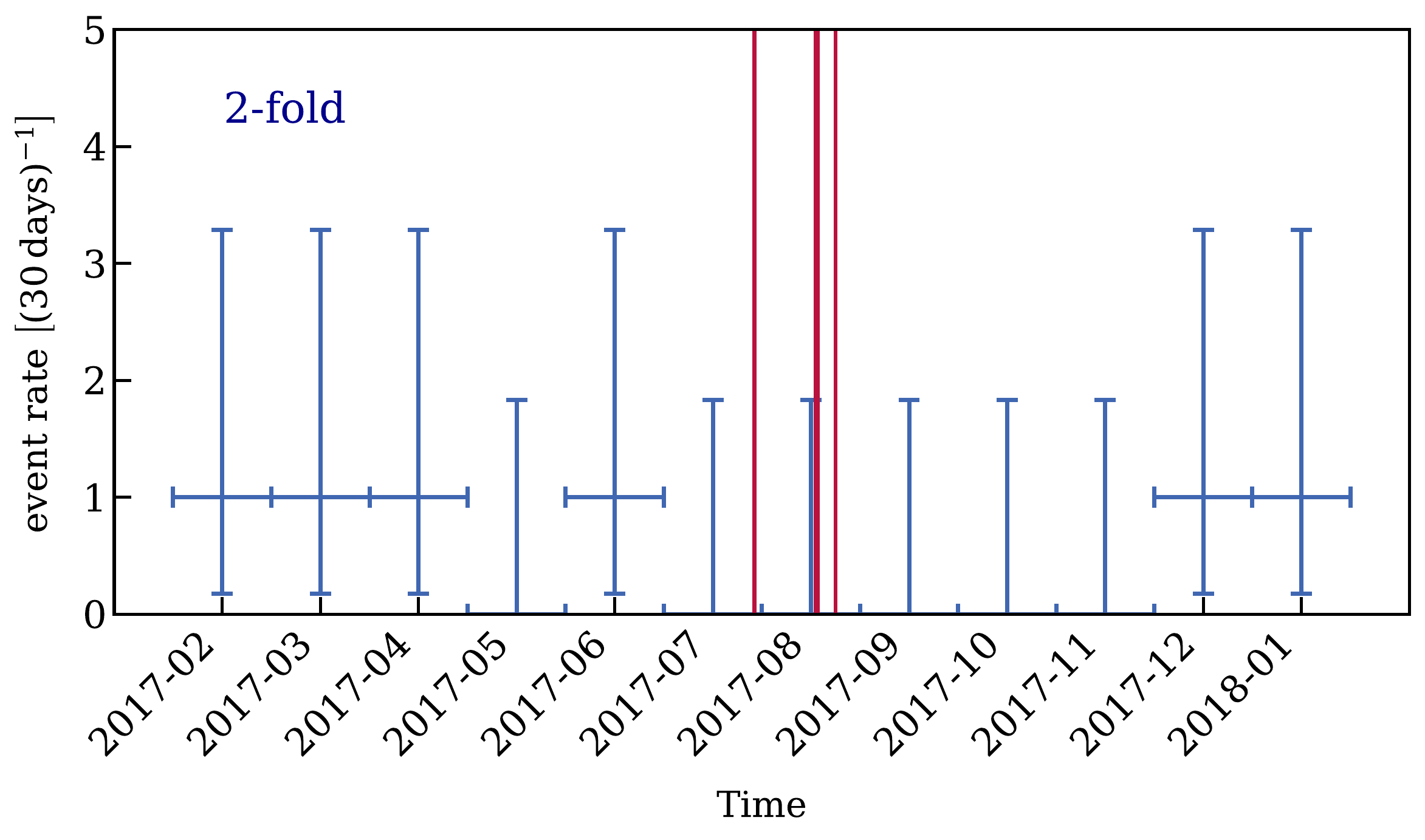}
\includegraphics[scale =0.34]{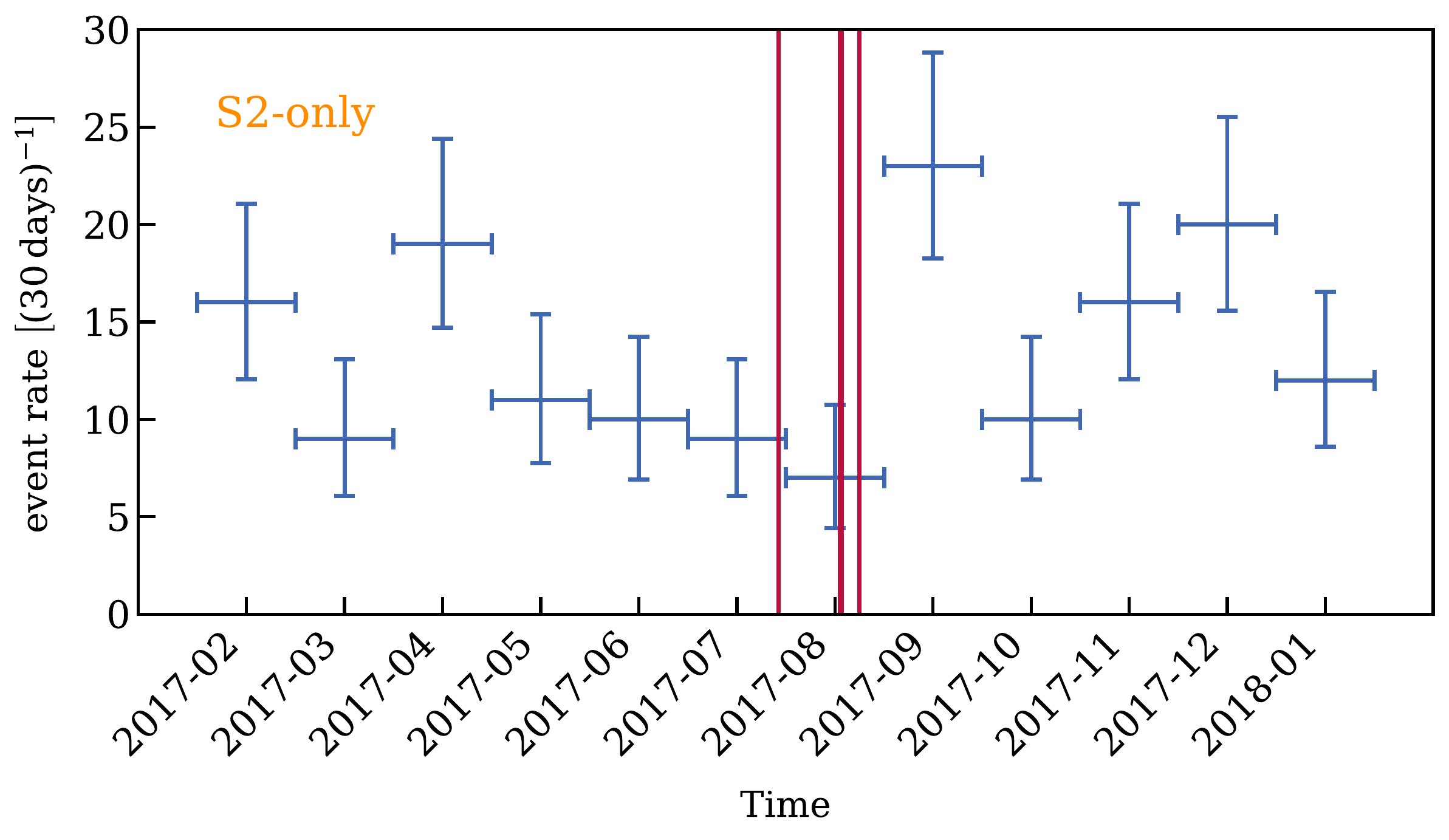}
\caption{\label{Event_distribution}The distribution of background events with time in ER and NR channels. 3-fold, 2-fold, and S2-only refer to different analysis modes of the NR channel. Red lines indicate the time of GW events. The rise and subsequent decay of event rate in the ER channel during March-September are due to production of $^{131\mathrm{m}}$Xe, $^{133}$Xe, and $^{125}$I during neutron calibration campaigns~\cite{2021seer.book.....S}.}
\end{figure*}

\begin{table*}[ht]
\caption{\label{tab:ER_Table} Predictions for the different energy bins of the ER channel during four GW events. For each bin, in the $\pm$\,500\,s time window around a GW event, we give the number of expected background events (N$_{\mathrm{exp}}$), the probability of observing zero background events (P(0)) and the minimum number of events corresponding to an excess at a 3\,$\sigma$ confidence level (N$_{\mathrm{th}}$).}
\setlength{\tabcolsep}{1pt}
\begin{ruledtabular}
\begin{tabular}{cccccccccccccccc
            }
&\multicolumn{3}{c}{\hspace{0.2cm}Low Energy} &\multicolumn{3}{c}{$^{83\mathrm{m}}$Kr} &\multicolumn{3}{c}{Medium Energy} &\multicolumn{3}{c}{\hspace{-0.6cm}$^{131\mathrm{m}}$Xe} &\multicolumn{3}{c}{\hspace{0.2cm}High Energy}  \\ \hline
& \hspace{0.5cm}N$_{\mathrm{exp}}$ & \hspace{-0.3cm}P(0) & \hspace{-0.6cm}N$_{\mathrm{th}}$ & \hspace{0.5cm}N$_{\mathrm{exp}}$ & \hspace{-0.4cm}P(0) & \hspace{-0.4cm}N$_{\mathrm{th}}$ & \hspace{0.6cm}N$_{\mathrm{exp}}$ & \hspace{-0.4cm}P(0) & \hspace{-1cm}N$_{\mathrm{th}}$ & N$_{\mathrm{exp}}$ & \hspace{-0.4cm}P(0) & \hspace{-0.4cm}N$_{\mathrm{th}}$ & \hspace{0.5cm}N$_{\mathrm{exp}}$& \hspace{-0.4cm}P(0) & \hspace{-0.6cm}N$_{\mathrm{th}}$ \rule{0pt}{2.5ex}\\ \hline

GW170729 & \hspace{0.5cm}0.04 & \hspace{-0.3cm}0.94& \hspace{-0.6cm}2 & \hspace{0.5cm}0.16 & \hspace{-0.4cm}0.85& \hspace{-0.4cm}3 & \hspace{0.5cm}0.22 & \hspace{-0.4cm}0.80 & \hspace{-1cm}3 & 0.64 & \hspace{-0.4cm}0.52 &\hspace{-0.4cm}5& \hspace{0.5cm}0.07& \hspace{-0.4cm}0.93 & \hspace{-0.8cm}2\rule{0pt}{2.5ex}\\
GW170817 & \hspace{0.5cm}0.04 & \hspace{-0.3cm}0.94& \hspace{-0.6cm}2 & \hspace{0.5cm}0.15 & \hspace{-0.4cm}0.86& \hspace{-0.4cm}3 & \hspace{0.5cm}0.22 & \hspace{-0.4cm}0.80 & \hspace{-1cm}3 & 0.29 & \hspace{-0.4cm}0.75 &\hspace{-0.4cm}4& \hspace{0.5cm}0.07& \hspace{-0.4cm}0.93 & \hspace{-0.8cm}2\\
GW170818 & \hspace{0.5cm}0.04 & \hspace{-0.3cm}0.94& \hspace{-0.6cm}2 & \hspace{0.5cm}0.15 & \hspace{-0.4cm}0.86& \hspace{-0.4cm}3 & \hspace{0.5cm}0.22 & \hspace{-0.4cm}0.80 & \hspace{-1cm}3 & 0.27 & \hspace{-0.4cm}0.76 &\hspace{-0.4cm}4& \hspace{0.5cm}0.07& \hspace{-0.4cm}0.93 & \hspace{-0.8cm}2\\
GW170823 & \hspace{0.5cm}0.04 & \hspace{-0.3cm}0.94& \hspace{-0.6cm}2 & \hspace{0.5cm}0.14 & \hspace{-0.4cm}0.87& \hspace{-0.4cm}3 & \hspace{0.5cm}0.22 & \hspace{-0.4cm}0.80 & \hspace{-1cm}3 & 0.23 & \hspace{-0.4cm}0.79 &\hspace{-0.4cm}3& \hspace{0.5cm}0.07 & \hspace{-0.4cm}0.93 & \hspace{-0.8cm}2\\
\end{tabular}
\end{ruledtabular}
\end{table*}
Although background events in these channels have been studied before, a direct comparison with the timing of gravitational waves has never been made. To ensure an unbiased analysis, we follow a blind approach where we refrain from utilizing the timing information of individual background events until the background model is finalized and the expected number of background events in the $\pm$500\,s time window is determined. 

Fig.\,\ref{Event_distribution} shows the distribution of background events with time in different analyses. Rates are constant within statistical uncertainty for the duration of the Science Runs for the NR channel. Therefore, for NR channel, the background rate is the ratio of the total number of background events during the entire Science Run and the total livetime of the entire Science Run. 

However, due to the background of activated isotopes --- $^{131\mathrm{m}}$Xe, $^{133}$Xe, and $^{125}$I; we need to account for the time-dependence of background rate in ER channel. Thus, we use the same background estimation for the ER channel as in~\cite{2021seer.book.....S}. 
Before unblinding the timing information of background events, we compare the values predicted by these background models with the actual observations in the different energy bins. The comparison was done using the $(-1500$,$-500)$\,s sideband before the GW events. To account for the degree of freedom due to five independent ER energy bins and four GW signals, we divided the chi-square by 20 and found the reduced chi-square statistic to be 1.14. This confirms that these background models agree with the observation.

Table~\ref{datasets} gives background rate and livetime in the units of events/day and seconds respectively. Therefore, the expected number of background events in the $\pm$\,$500$\,s time window around a GW signal, is calculated as,
\begin{equation}
    \mathrm{N}_{\mathrm{exp}} = \mathrm{background\:rate}\times\frac{\mathrm{livetime}}{24\times60\times60}.
\end{equation}
The corresponding probabilities of observing zero background events are values of cumulative distribution functions at zero, assuming Poisson statistics. These values are given in Tab.\,\ref{tab:Channels} for the NR channel and in Tab.\,\ref{tab:ER_Table} for the ER channel. 

The $\pm$500\,s time window was blinded during background estimation, and we calculated the minimum number of events in each bin of the ER channel for each GW, above which evidence for a particle signal in that bin could have been claimed, using the expected number of background events, based on the background model. A potential excess at a 3\,$\sigma$ confidence level corresponds to a p-value of 3\,$\times$\,$10^{-3}$. However, this local p-value needs to be corrected to account for the look-elsewhere effect due to 20 independent bins. The look-elsewhere effect refers to an increased chance of a false positive for a larger number of trials due to statistical fluctuations~\cite{Bayer:2020pva}. To correct the conclusions of a hypothesis test, Sidak correction defines the global p-value for $n$ independent bins as~\cite{doi:10.1080/01621459.1967.10482935},
\begin{equation}
    \mathrm{p}_{\mathrm{global}} = 1-(1-\mathrm{p}_{\mathrm{local}})^{1/n}.
\end{equation}
The correction yields a global p-value of 1.5\,$\times$\,$10^{-4}$ for a potential excess at a 3\,$\sigma$ confidence level. The corresponding minimum numbers of required events for reporting excess, calculated using the expected number of background events following Poisson distribution, are between 2 and 5, as shown in Tab.\,\ref{tab:ER_Table}. In the NR channel, values of cumulative distribution functions at 1, for Poisson distributions with such small means, are more than 99.99\%; implying that even 1 event corresponds to an excess.

\section{Results}\label{results}

\begin{figure*}[!htbp]
\includegraphics[scale =0.5]{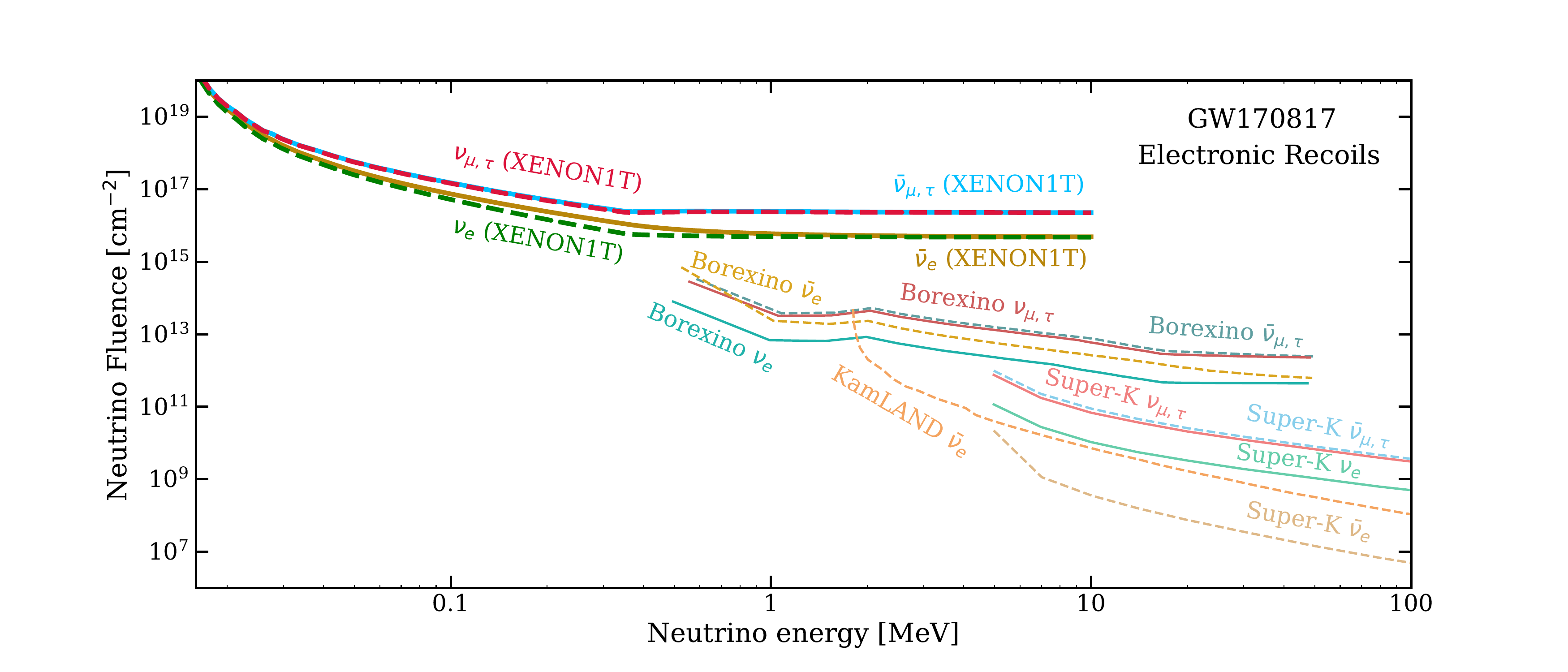}
\caption{\label{limits-CC} The upper limit on the fluence (time-integrated flux) of coincident neutrinos emitted in GW170817, the merger of neutron stars using the electronic recoil signals in XENON1T, obtained for electron neutrinos ($\nu_e$, dashed green), electron antineutrinos ($\bar{\nu}_e$, solid gold), muon or tau neutrinos (${\nu}_{\mu, \tau}$, dashed red), and muon or tau antineutrinos ($\bar{\nu}_{\mu, \tau}$, solid cyan). We also show the limits obtained by Borexino~\cite{BOREXINO:2023nji}, KamLAND~\cite{KamLAND:2020ses}, and Super-K~\cite{Super-Kamiokande:2018dbf} for comparison.} 
\end{figure*}
Once these detection channels, time and energy ranges, and background models were fixed, the timing information of background events was unblinded. No background event was observed in any of these channels within $\pm$\,500 seconds of any tested GW signal. Given the low background rates, this is expected under the background-only hypothesis in the NR channel, as can be seen in Tab~\ref{datasets}. The total number of expected background events in all energy bins of the ER channel across all GWs is $\sim$3. However, even in this channel, no background events were observed, consistent with the background-only hypothesis, assuming Poisson statistics, within a 3\,$\sigma$ confidence interval.

\begin{figure*}[!htbp]
\includegraphics[scale =0.5]{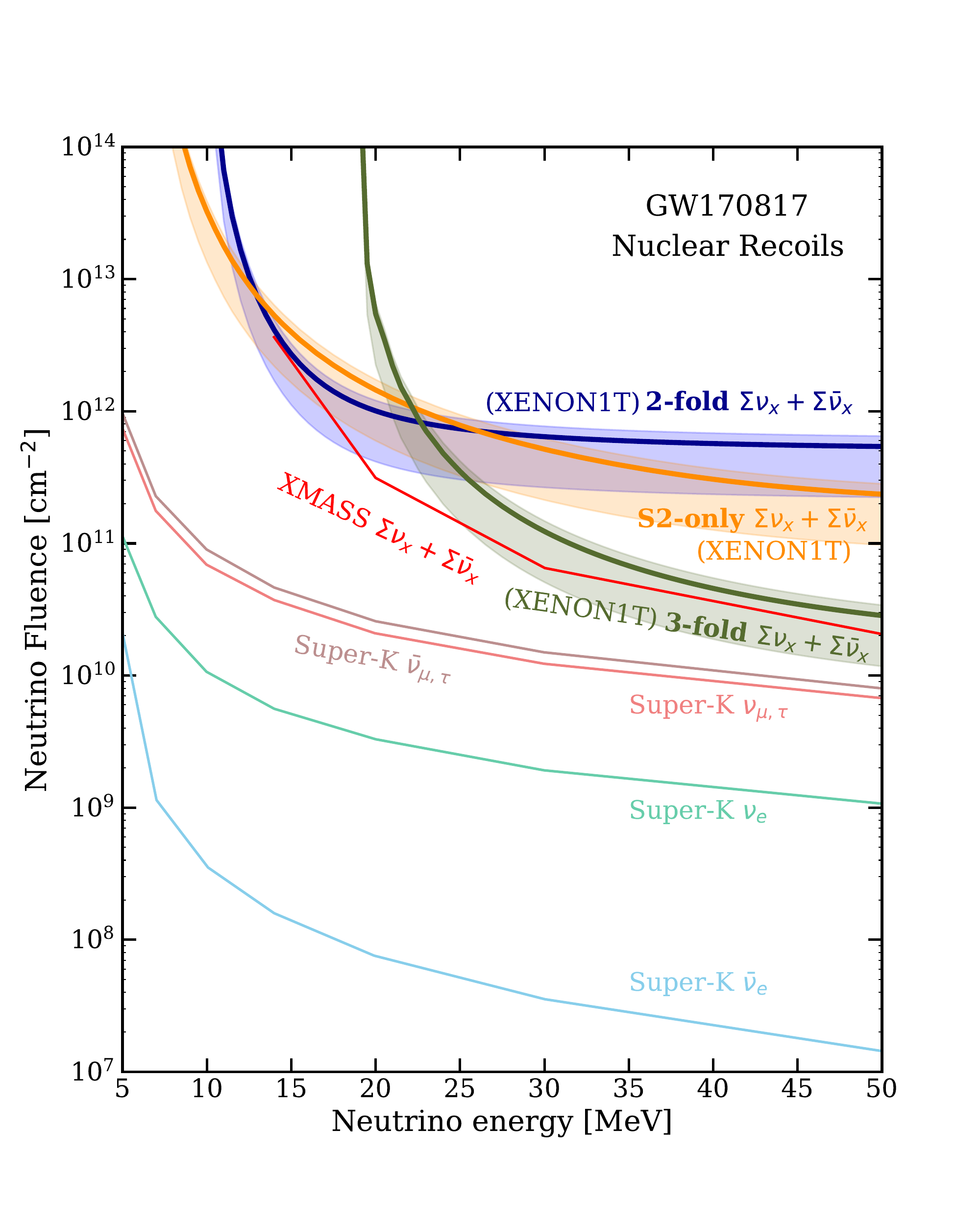}
\caption{\label{limits-CEvNS} The upper limit on the fluence (time-integrated flux) of coincident neutrinos emitted in GW170817, the merger of neutron stars using the nuclear recoil signals in XENON1T, placed on the sum of the fluence of all flavors of neutrinos and antineutrinos. 3-fold (olive-green), 2-fold (blue), and S2-only (orange) refer to different modes in which the data was analyzed in the search for nuclear recoils, as described in Sec.\,\ref{datasets}. We also show the limits obtained by XMASS~\cite{XMASS:2020jdj} and Super-K~\cite{Super-Kamiokande:2018dbf} for comparison.} 
\end{figure*}

The absence of background events can be converted into an upper limit on coincident neutrinos emitted in the mergers, assuming a Poisson distribution of  neutrino signal events~\cite{Feldman:1997qc}. To ensure consistency with other experiments, we place limits at a 90\% confidence level on fluence, which represents the integration of flux with respect to time. As explained in section~\ref{datasets}, these limits are on the fluence from the merger of neutron stars -- GW170817. The upper limit on neutrino fluence at 90\% confidence level using ER channel is shown in Fig.\,\ref{limits-CC}. As can be seen, we place upper limits on the fluence of $\mathcal{O}$(10 keV) neutrinos. No other detector has been able to probe such low-energy neutrinos. The upper limit on neutrino fluence at 90\% confidence level using NR channel is shown in Fig.\,\ref{limits-CEvNS}. Our limits are comparable to those from the XMASS dark matter detector. Both limits are on the sum of the fluence of all flavors of neutrinos and antineutrinos, which is different for the stronger limits from Super-K. Further, all the limits (Fig.\,\ref{limits-CC} and \ref{limits-CEvNS}) correspond to neutrino fluxes higher than the typical mass of the binary systems for isotropic emission.

We also use our data to constrain the fluence of axion-like particles (ALPs) with keV energies coincident with GW events. In XENON1T, ALPs can be detected via the axio-electric effect, a process analogous to the photoelectric effect~\cite{Derevianko:2010kz}. Absorption of an axion causes atomic ionization, and the energy is transferred from the axion to the electron during the interaction. Similar to neutrinos, we assume that ALPs are emitted mono-energetically from mergers. Therefore the number of ERs in the detector due to ALPs can be expressed as,
\begin{equation}
    N_\mathrm{ER} = N_T\,\epsilon\,\sigma_\mathrm{ALP}\, F_{E_\mathrm{ALP}},
\end{equation}
where $\sigma_\mathrm{ALP}$ is the cross section of ALPs to interact with electrons and $F_{E_\mathrm{ALP}}$ is the fluence of ALPs with energy $E_\mathrm{ALP}$. It can be seen in Fig.\,2 of~\cite{XENON:2020rca} that the efficiency of our detector, $\epsilon$ for the ER channel is $\sim$89\% for recoil energies greater than 5~keV$_{\mathrm{ee}}$. The number of electrons available per atom when the recoil energy is greater than 34.6 (5.5) keV$_{\mathrm{ee}}$ is 54 (52) electrons. Given that we do not see any background event, we obtain the upper limit on the product of coincident fluence and cross section: $\sigma_\mathrm{ALP} F_\mathrm{E_{ALP}} < 10^{-29}$~cm$^2$/cm$^2$ for $E_\mathrm{ALP}$ in [5.5--210]\,keV$_{\mathrm{ee}}$. 

\section{Conclusions}
We have provided the first search for particle signals in the keV energy region associated with GWs. Despite the 1.3 tonnes fiducial mass of XENON1T, various analysis modes and energy regions down to 1 keV$_{\mathrm{ee}}$, no particle signals were found within $\pm$500\,s time windows around GW signals.

As a result, we extended the upper limits on the coincident neutrino fluence down to neutrino energies of 17~keV. Our limits (Fig.\,\ref{limits-CC} and \ref{limits-CEvNS}) on neutrino fluence correspond to values higher than the typical mass of the binary systems for isotropic emission at 40 Mpc (the closest merger). These limits might be constraining if the merger is at distances less than 10 kpc. Therefore, the current and next generation of xenon detectors~\cite{XENON:2023sxq,LZ:2022ufs,Liu:2022zgu,Aalbers:2022dzr,angelides} could probe meaningful parameter space for gravitational waves from Galactic mergers. However, such a galactic merger would be rare -- once in 100 centuries ~\cite{Olejak:2019pln}. 

We put an upper limit on the product of coincident fluence and cross section of ALPs in the [5.5--210]\,keV energy range interacting with atomic electrons. Existing models discuss the emission of axions at the MeV level, but no theoretical models exist for the emission of axions at the keV level. Nevertheless, these limits may help motivate models of binary star mergers involving Beyond Standard Model particles and prove a useful test of future models.
\section{Acknowledgements}

We gratefully acknowledge support from the National Science Foundation, Swiss National Science Foundation, German Ministry for Education and Research, Max Planck Gesellschaft, Deutsche Forschungsgemeinschaft, Helmholtz Association, Dutch Research Council (NWO), Weizmann Institute of Science, Israeli Science Foundation, Binational Science Foundation, R\'egion des Pays de la Loire, Knut and Alice Wallenberg Foundation, Kavli Foundation, JSPS Kakenhi and JST FOREST Program in Japan, Tsinghua University Initiative Scientific Research Program and Istituto Nazionale di Fisica Nucleare. This project has received funding/support from the European Union’s Horizon 2020 research and innovation programme under the Marie Sk\l{}odowska-Curie grant agreement No 860881-HIDDeN. Data processing is performed using infrastructures from the Open Science Grid, the European Grid Initiative and the Dutch national e-infrastructure with the support of SURF Cooperative. We are grateful to Laboratori Nazionali del Gran Sasso for hosting and supporting the XENON project.
\newpage

\bibliographystyle{apsrev}
\bibliography{paper} 

\end{document}